\documentstyle[12pt,a4]{article}
\input{psfig.sty}
\input{amssymb.sty}

\def\bbc{{\mathbb C}}

\def\bbr{{\mathbb R}}

\def\bbz{{\mathbb Z}}

\def\tr{\mathop{\rm tr}\nolimits}

\def\frac#1#2{{{#1}\over{#2}}}

\begin{document}
\begin{center}
{\large\bf Some uses of moduli spaces in particle and field theory}\\
TSOU Sheung Tsun\\
Mathematical Institute, Oxford University\\
24--29 St.\ Giles', Oxford OX1 3LB\\
United Kingdom.\\
tsou\,@\,maths.ox.ac.uk
\end{center}

{\bf Abstract}\\

In this talk I shall try to give an elementary introduction to certain
areas of mathematical physics where the idea of moduli space is used
to help solve problems or to further our understanding.  In the wide
area of gauge theory, I shall mention instantons, monopoles and
duality.  Then, under the general heading of string theory, I shall
indicate briefly the use of moduli space in conformal field theory and
$M$-theory.

\vfill
\noindent{Invited talk at the EWM Workshop on Moduli Spaces, 2--3
July, Oxford; Proceedings edited by Frances Kirwan, Sylvie Paycha,
Tsou Sheung Tsun, produced by Mathematical Institute, Oxford University.}

\section{Introduction}
Physicists seldom define their terms.  So although I know roughly what
a moduli space is, and the sort of thing one does with it in physics,
I was not really very sure of what exactly it is.  So I asked Frances
(Kirwan), just as the Balliol College (where participants were lodged)
porters did when they also 
wanted to know what a
moduli space was.  I have always taken it to be some sort of useful
parameter space, convenient in the sense that mathematicians have
already worked out all its properties (at least in the classical
cases).  But Frances told me something much more significant---she
describes it as a parameter space in the {\em nicest possible way}.

So in the next 55 minutes or so, I shall try to give you a rough
picture of how physicists have made use of this nice concept of a
parameter space.  We should note, however, that it is far from a one-way
traffic.  Much of the tremendous progress in 4-manifold theory, and a
large part of it is done here, came about by studying certain moduli
spaces occurring in mathematical physics.

A few notes of warning, however, are in place.  For a hard-nosed or
pragmatic physicist, (A) spacetime $X$ has 4 dimensions, 3 space and 1
time, with an indefinite metric.  By an indefinite metric I mean that
the quadratic form giving the metric is not positive definite, so that
two distinct points in spacetime can be null-separated.  In fact,
distances along light-paths are always zero.  For him (or her) also
(B) spacetime is by and large like $\bbr^4$, that is, (i) flat, (ii)
looking more or less the same in all directions, (iii) real, and (iv)
more or less infinite in all its 4 directions and hence non-compact.

On the other hand, algebraic geometry is more about Riemannian
manifolds and the best results are almost always obtained for the
compact case.  In order to make contact, the concept of spacetime has
to be modified in several significant ways.
\begin{enumerate}
\item One considers definite metrics, a process known as {\em
euclideanization}.  Then many nice things happen.   In particular, the
{\em wave operator}
$$ \Box = \frac{\partial^2}{\partial t^2} -\frac{\partial^2}{\partial
x^2}  - \frac{\partial^2}{\partial y^2} - \frac{\partial^2}{\partial
z^2}  $$
which is hyperbolic, becomes the 4-dimensional Laplacian
$$ \nabla^2 = \frac{\partial^2}{\partial t^2} +\frac{\partial^2}{\partial
x^2}  + \frac{\partial^2}{\partial y^2} + \frac{\partial^2}{\partial
z^2}  $$
which is elliptic, and for elliptic operators there are all sorts of
good results like the index theorems.   Euclideanization is done in
the following: Self-dual Yang--Mills theory, instantons, monopoles,
Seiberg--Witten theory, strings, \ldots .
\item Alternatively, one {\em complexifies} spacetime, and then the
question of definite or indefinite metric disappears.  In this case,
one can use powerful complex manifold techniques including twistor
theory.  This is also where supersymmetry comes in mathematically.
Moreoever, by a change of point of view (see later), Riemann surfaces
also play an important role.   Complexification is done in
superstrings, supersymmetric Yang--Mills theory, $M$-theory, \ldots .
\item One also changes the topology of spacetime by {\em
compactifying} some or all of its directions.  In some cases, this is
only a mild change, amounting to imposing certain decay properties at
infinity (see later).  In other cases, this gives rise to important
symmetries of the theory.   Compactification is done in instantons,
superstrings, $M$-theory, \ldots .
\item One either changes the number of spacetime dimensions or
re-interprets some of them as other degrees of freedom.  This
dimensional change is done in strings, superstrings, monopoles,
$M$-theory, \ldots .
\end{enumerate}

At first sight, these modifications look drastic.  The hope is that
they somehow reflect important properties of the real physical world,
and that the nice results we have do not disappear on us once we know
how to {\em undo} the modifications.  Surprisingly, the (largely
unknown) mathematics underlying real 4-dimensional spacetime looks at
present quite intractable!

\section{Yang--Mills theory (Gauge theory)}
Unlike most of the other theories I shall mention, Yang--Mills theory
is an experimentally `proven' theory.  In fact, it is generally
believed, even by hard-nosed or pragmatic physicists, that Yang--Mills
theory is the basis of {\em all} of particle physics.  From the
physics point of view, Yang--Mills theory is the correct framework to 
encode the invariance of particle theory under the action of a
symmetry group---the gauge group $G$---at each spacetime point.  For
example, let $\psi (x)$ be the wave-function of a quantum particle.
Then the physical system is invariant under the action of the group:
$$ \psi(x) \mapsto \Lambda (x) \psi(x), \quad \Lambda (x) \in G.  $$
This invariance is known as {\em gauge invariance}.
Now the groups that are most relevant to particle physics are $U(1),
SU(2), SU(3)$.   However, we shall come across other groups as well.
But for simplicity, we shall take $G=SU(2)$, unless otherwise stated.

There is an additional ingredient in many favoured gauge theories,
namely {\em supersymmetry}.  This is a symmetry relating two kinds of
particles: bosons (e.g.\ a photon) with integral spin and fermions
(e.g.\ an electron) with half-integral spin.  Spin is a kind of
internal angular momentum which is inherently quantum mechanical.
Since bosons and fermions in general behave quite differently (e.g.\
they obey different statistics), this symmetry is not observed in
nature.  However, one can imagine this symmetry holding for example at
ultra-high energies.  What makes this symmetry theoretically
interesting is that many theories simplify and often become complex
analytic with this extra symmetry, making much of the underlying
mathematics accessible.  Also the complex analyticity links such
theories with most studies of moduli spaces.

\begin{figure} [h]
\centering
\centerline{\psfig{figure=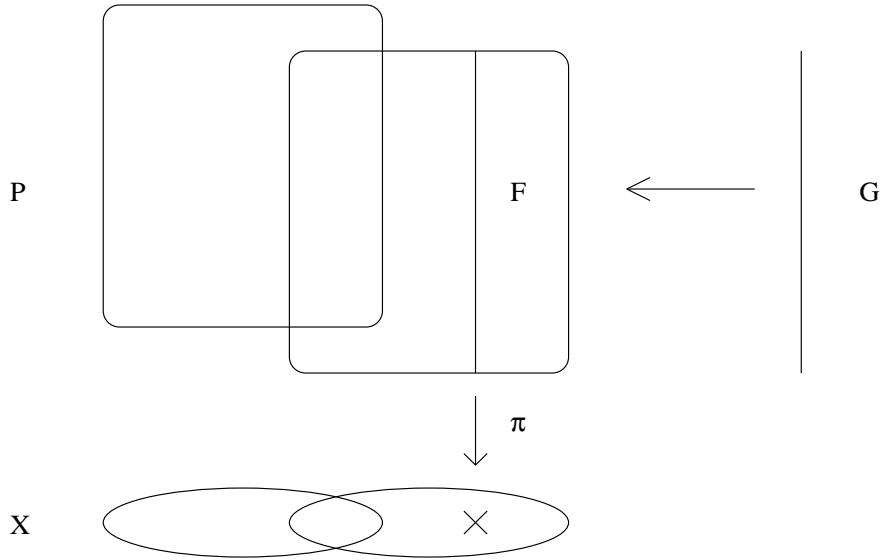,width=0.8\textwidth}}
\caption{Sketch of a principal bundle}
\label{pbundle}
\end{figure}

Mathematically, Yang--Mills theory can be modelled (in the simplest
case) by a principal bundle $P$ (see Figure \ref{pbundle})
together with a connection on it.  I remind you that, roughly
speaking, a {\em principal bundle} is a manifold $P$ with a projection $\pi$
onto a {\em base space} $X$, and a right action by the {\em structure 
group} $G$.  In general, the base space can be any smooth manifold,
but here we consider only the case of spacetime $X$.
Above each point $x \in X$, the inverse image (called the
{\em fibre}) $\pi^{-1} (x)$ is homeomorphic to $G$.  The {\em total
space} $P$ is locally a product, in the sense that $X$ is covered by
open set $U_\alpha$ and $\pi^{-1} (U_\alpha)$ is homeomorphic to
$U_\alpha \times G$.  A {\em connection} $A$ 
is a 1-form on $P$ with
values in the Lie algebra $\frak{g}$ of $G$, satisfying certain
conditions and giving a prescription for differentiating vectors and
tensors on $X$.  It combines with the usual exterior derivative $d$ to
give the covariant exterior derivative $d_A$:
$$ d_A = d + A $$
in such a way as to preserve gauge invariance.

Next we need the curvature 2-form:
$$ F_A=dA+A A \quad (F_{\mu\nu}=\partial_\nu A_\mu - \partial_\mu A_\nu
+ig [A_\mu,A_\nu]).  $$
The second formula (in brackets) is the same as the first one, but
written in local coordinates, or `with indices', where $\mu=0,1,2,3.$

Since $\dim X=4$ (for the moment, anyway), we have the Hodge star
operator which takes 2-forms to 2-forms:
\begin{eqnarray*}
 * \colon & \Omega^2 & \to \Omega^2\\
 & F_A & \mapsto {}^*\!F_A.
\end{eqnarray*}
In local coordinates, this can be written as
$$ {}^*\!F_{\mu\nu}=-\frac{1}{2} \epsilon_{\mu\nu\rho\sigma}
F^{\rho\sigma},$$
where $\epsilon_{\mu\nu\rho\sigma}$ is a completely skew symbol
defined by $\epsilon_{0123}=1$.
Notice that
\begin{eqnarray*}
 (*)^2 & = &+1 \quad {\rm in\ euclidean\ metric}\\
 (*)^2 & = &-1 \quad {\rm in\ Minkowskian\ metric.}
\end{eqnarray*} 

Yang--Mills theory is given by the Yang--Mills action or functional
$$S(A) = \frac{1}{8 \pi^2} \int_X \tr (F_A {}^*\!F_A) =\frac{1}{8
\pi^2} \|F_A\|^2.  $$
The curvature satisfies:
\begin{eqnarray*}
d_A F_A & = & 0 \quad {\rm (Bianchi\ identity)}\\
d_A {}^*\!F_A &  = & 0 \quad {\rm (Yang\!-\!Mills\ equation).}
\end{eqnarray*}
These are the {\em classical} equations for Yang--Mills theory.
Notice that the first one is an identity from differential geometry, 
and the second one comes
from the first variation of the action.

The space of connections $\cal A$ is an affine space, but we are
really interested in connections modulo gauge equivalence.  Two
connections $A,A'$ are gauge equivalent if they are `gauge transforms'
of each other:
$$ A'= \Lambda^{-1} A \Lambda + \Lambda^{-1} d \Lambda. $$
In other words, $\Lambda (x) \in G$, $\Lambda$ is a
fibre-preserving automorphism of $P$ invariant under the action of
$G$.  We shall use the symbol $\cal G$ for the group of gauge
transformations $\Lambda$.

So we come to our first, most basic, {\em moduli space}
$$ \bar{\cal M} = {\cal A}/{\cal G}.  $$
It is in general infinite-dimensional with complicated topology.

We shall be interested in various subspaces or refinements of
$\bar{\cal M}$.

One theoretical use of $\bar{\cal M}$ itself is in (the euclidean
formulation of) quantum field theory, where with the Feynman path
integral approach, one has to consider the integral of the exponential
of the
Yang--Mills
action over $\bar{\cal M}$:
$$ \int_{\bar{\cal M}} e^{-S(A)}.  $$
But this integral is very difficult to define in general!

The moduli space $\bar{\cal M}$ has a singular set which represents
the {\em reducible connections}, which are connections with holonomy
group $H \subset G$ such that the centralizer of $H$ properly contains
the centre of $G$.  We say then that the connection {\em reduces} to $H$.
The complement $\cal M$ of this singular set is
dense in $\bar{\cal M}$, and represents the irreducible connections.
For $G=SU(2)$, near an irreducible connection $\bar{\cal M}$ is
smooth, but reducible connections lead to cone-like singularities in
$\bar{\cal M}$. 
 
\subsection{Instantons}
Recall that $G=SU(2)$.  Bundles $P$ over $X$ are classified by the
second Chern class of the associated rank 2 vector bundle $E$ (cf.\
Rosa-Maria Mir\'o-Roig's talk):
$$k=c_2(E)[X] = \frac{1}{8 \pi^2} \int_X \tr F_A^2 \in \bbz. $$

We say that a connection $A$ is self-dual (or anti-self-dual) if its
curvature $F_A$ satisfies
$$ F_A={}^*\!F_A \quad ({\rm resp.}\ F_A=-{}^*\!F_A).  $$
Then given any connection $A$, we can decompose the corresponding  
curvature $F_A$ into its self-dual and and anti-self-dual parts: 
$$ F_A=F_A^+ +F_A^-.  $$

In the context of Yang--Mills theory a self-dual connection is called
an {\em instanton}\footnote{It is a matter of convention whether one
so defines a self-dual or anti-self-dual connection.}:
$$ F_A={}^*\!F_A \Leftrightarrow F_A^-=0. $$
In this case, 
$$ {\rm Bianchi\ identity} \cong {\rm Yang\!-\!Mills\ equation}.  $$
In other words, a self-dual connection is {\em automatically} a
classical solution.

Now we have
\begin{eqnarray*}
S(A) & = & \frac{1}{8 \pi^2} \int_X |F_A^+|^2 + |F_A^-|^2 \\
k & = & \frac{1}{8 \pi^2} \int_X |F_A^+|^2 - |F_A^-|^2.
\end{eqnarray*}
Hence one has immediately
$$S(A) \geq k, $$
and
$$ S(A) =k \Leftrightarrow F_A^-=0.  $$
So a self-dual connection gives an absolute minimum for the action.
The integer $k$ is known as the {\em instanton number}.

\underline{Warning:} Nontrivial self-dual connections exist only when
$X$ is either euclidean or complex.

The {\em mathematical} magic of instantons is that instead of
solving the second order Yang--Mills equations we have only the first
order self-duality equation to deal with.  These connections can
actually be constructed using euclidean {\em twistor methods} without
explicitly solving any equations (cf.\ Tatiana Ivanova's talk).

\begin{figure} [h]
\centerline{\psfig{figure=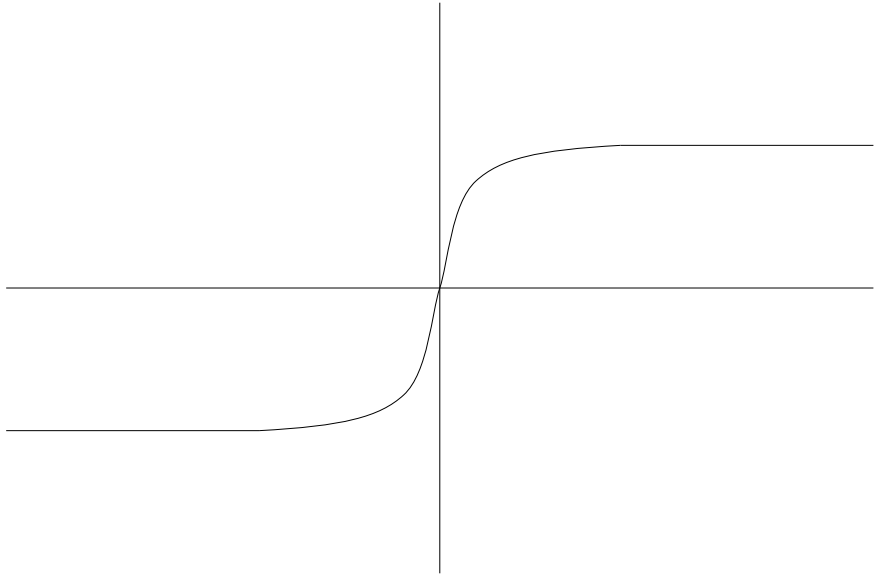,width=0.8\textwidth}}
\caption{Sketch of a kink connecting two different states}
\label{kinks}
\end{figure}

{\em Physically}, the presence of instanton contribution in the
path integral allows tunnelling between different vacua (i.e.\ lowest
energy states) of the relevant Yang--Mills theory (namely quantum
chromodynamics for strong interactions or
QCD).  This
role of the instantons can be compared to lower-dimensional objects
such as `solitons' or topological defects called `kinks' which connect
up two different states at infinity (see Figure \ref{kinks}).  The two
phenomena are quite similar, since `tunnelling' means a quantum
particle can penetrate a potential barrier which a classical particle
cannot go through, thus connecting two classically separate states.
The effect of instantons is `non-perturbative' in the sense that such
an effect cannot be obtained as a term in a power series expansion of
$g$ the coupling constant (which is measure of the `strength' of the
interaction under consideration, and which appears for example in the
nonlinear term of the curvature form $F_{\mu\nu}$).  This is a 
direct manifestation of
the fact that instantons are topological in nature and cannot be
obtained by any `local' considerations such as power series
expansions.

Since in euclidean space the Yang--Mills equations are elliptic, and
concentrating on irreducible connections gets rid of zero eigenvalues,
one can use the index theorem to count the `formal dimension' of {\em
instanton moduli space}.  Typically the smooth part of the moduli
space will have this formal dimension as its actual dimension.  For
example, 
$$ X=S^4, \quad \dim_\bbc({\cal M}_{I,k}) = 8k-3.  $$
Uhlenbeck has given a unique compactification of ${\cal M}_I$, the
union for all $k$.
For more details about instanton moduli spaces, I again refer you to
Tatiana Ivanova's talk.

\subsection{Monopoles}
Recall $G=SU(2)$.

Consider a Yang--Mills theory with a scalar field (called {\em Higgs
field}) $\phi$, together with a potential term $V(\phi)$ which is
added to the Yang--Mills action.  Suppose further that
$$ V(\phi_0)= {\rm minimum\ for}\ |\phi_0| \neq 0,  $$
and that $V(\phi)$ is invariant under a subgroup $U(1) \subset
SU(2)$.  Then for those connections of $P$ which are reducible to this
$U(1)$ subgroup, we can for certain purposes concentrate on this
`residual gauge symmetry' and have a $U(1)$ gauge theory.  If we
interpret this $U(1)$ as Maxwell's theory of electromagnetism, then a
non-trivial reduction of $P$ can be regarded as a {\em magnetic
monopole}.  The {\em magnetic charge} $k$ is given by the first Chern
class of the reduced bundle.  In fact we have the following exact
sequence which gives us an isomorphism:
$$
\begin{array}{ccccccc}
\pi_2 (SU(2)) & \rightarrow & \pi_2(SU(2)/U(1)) & \stackrel{\sim}{
\rightarrow} &
\pi_1 (U(1)) & \rightarrow & \pi_1 (SU(2)) \\
\parallel &&&&&&\parallel \\
0 &&&&&& 0
\end{array}  $$

  Unlike the original magnetic monopole considered by Dirac, these
't~Hooft--Polyakov monopoles have finite energy and are the soliton
solutions of the field equations corresponding to the action:
$$ S(A,\phi)= S(A) + \|D\phi\|^2 + \lambda (1-|\phi|^2)^2,  $$
where the last term is the usual form of the potential $V(\phi)$.
From this we get the Yang--Mills--Higgs equations (YMH):
\begin{eqnarray*}
D_A F & = & 0,\\
D_A {}^*\!F & = & -[\phi,D_A \phi],\\
D_A {}^*\!D_A \phi & = & 2 \lambda \phi (|\phi|^2 -1).
\end{eqnarray*}

Now we specialize to a certain limit, the Prasad-Somerfeld limit:
$V(\phi)=0$, but $|\phi| \to 1$ at infinity.  Then the
Yang--Mills--Higgs system becomes:
\begin{eqnarray*}
D_A F & = & 0,\\
D_A {}^*\!F & = & -[\phi,D_A \phi],\\
D_A {}^*\!D_A \phi & = & 0.
\end{eqnarray*}

Consider next a Yang--Mills theory in euclidean $\bbr^4$, invariant under
$x_4$-translations.  Then we can write
$$ A=A_1 dx_1 + A_2 dx_2 + A_3 dx_3 + \phi dx_4,  $$
where $A_1,A_2,A_3,\phi$ are Lie algebra-valued functions on
$\bbr^3$.  The action can be written as
$$S(A)= \|F_A\|^2 = \|F\|^2 + \|D \phi\|^2,  $$
where now  $F$ is the curvature of the connections in 3 dimensions:
$$A'=A_1 dx_1 + A_2 dx_2 + A_3 dx_3, $$
and $D$ is the corresponding 3-dimensional covariant derivative.
In this way, we can make the following identification since the
actions for the two theories are identical:
$$ {\rm YMH\ on}\ \bbr^3 \cong {\rm dimensionally\ reduced\ YM\ on}\
\bbr^4. $$
In this case, 
$$ F_A = {}^*\!F_A \Rightarrow {\rm first\ 2\ YMH}.  $$
Hence a solution to the Bogomolny equation
$$ F= {}^*\!D_A \phi $$
gives a solution of YMH.  These are known as `static monopoles'.

The moduli spaces ${\cal M}_k$ corresponding to a given charge $k$ are
well studied, at least
for $k=1,2$.  The translation group $\bbr^3$ acts freely on ${\cal
M}_k$, so does an overall phase factor $S^1$.  Dividing these out we
get the reduced monopole moduli spaces ${\cal M}_k^0,\
\dim_\bbc=4k-4$.  Taking the $k$-fold covers, one obtains:
$$ \tilde{\cal M}_k \cong \bbr^3 \times S^1 \times \tilde{\cal
M}_k^0.  $$

The special case of $k=2$ has been studied by Atiyah and Hitchin as an
entirely novel way of obtaining the scattering properties of two
monopoles, using a metric on ${\cal M}_2^0$ they discovered, and
assuming (with Manton) that geodesic motion on it describes adiabatic
motion of the two monopoles.  This is the most direct use that I know
of of moduli space for deriving something akin to dynamics!

\subsection{Topological field theory}
I wish just to mention a class of quantum field theories called {\em
topological quantum field theories} (TQFT), where the observables
(correlation functions) depend only on the global features of the
space on which these theories are defined, and are independent of the
metric (which, however, may appear in the classical theory).  Atiyah
gave an axiomatic approach to these, but there are so many local
experts here that I do not feel justified in expanding on that!

Instead, I shall just indicate the role of moduli space in Witten's
approach.  Starting with a moduli space $\cal M$ one can get fields,
equations and symmetries of the theory.  Witten postulates the
existence of certain operators ${\cal O}_i$ corresponding to
cohomology classes $\eta_i$ of $\cal M$ such that
$$ \langle {\cal O}_1 \cdots {\cal O}_n \rangle = \int_{\cal M} \eta_1
\cdots \eta_n,   $$
where $\langle \cdots \rangle$ denotes the correlation function of the
operators.  Hence he obtains these correlation functions as
intersection numbers of $\cal M$, using Donaldson theory.  So in a
sense the TQFT is entirely defined by $\cal M$.

The observables called correlation functions can best be understood in
the case of, for example, a 2-point function in statistical
mechanics.  This is the probability, given particle 1, of finding 
particle 2 at another fixed location.

To go into any further details about TQFT would require 
more detailed knowledge both of quantum field theory and
supersymmetry.  These would lead us unfortunately too far from the 
context of this workshop.

\subsection{Seiberg--Witten theory}
Recall that a spin structure on $X$ is a lift of the structure group
of the tangent bundle of $X$ from $SO(4)$ to its double cover
Spin(4)$\,\cong\,SU(2) \times SU(2)$.  Because of this isomorphism,
one can represent a spin structure more concretely as a pair of
complex 2-plane bundles $S^+,S^- \to X$, each with structure group
$SU(2)$.  A slightly more general concept is a
spin$^c$ structure over $X$, which is given by a pair of vector
bundles $W^+,W^-$ over $X$ with an isomorphism for the second exterior
powers
$$\Lambda^2 W^+ = \Lambda^2 W^- = L,\ {\rm say}, $$
such that one has locally
$$ W^\pm = S^1 \otimes L^{\frac{1}{2}}, $$
where $L^{\frac{1}{2}}$ is a local square root of $L \colon
L^{\frac{1}{2}} \otimes L^{\frac{1}{2}} = L$.

Given a spin$^c$ manifold $X$,
the Seiberg--Witten equations (SW)  are written for a system 
consisting of 1) a
unitary connection $A$ on $L=\Lambda^2 W^\pm$, and 2) $\psi$ a section
of $W^+$.  Then these equations are:
\begin{eqnarray*}
D_A \psi & = &0\\
F_A^+ & = & -\tau (\psi,\psi),
\end{eqnarray*}
where $\tau$ is a sesquilinear map $\tau \colon W^+ \times W^+ \to
\Lambda^+ \otimes \bbc$.

The Seiberg--Witten equations (SW) can be obtained from varying the
following functional:
$$ E(A,\psi) = \int_X |D_A \psi|^2 + |F_A^+ + \tau(\psi,\psi)|^2 +
R^2/8 + 2\pi^2 c_1 (L)^2,  $$
where $R$ is the scalar curvature of $X$ and $c_1(L)$ is the first
Chern class of $L$.  Notice that the last two terms depend only on $X$
and $L$, so that solutions of SW are absolute minima of $E$ on the
given bundle $L$.

The relevant moduli space here is the space $\cal M$ of all
irreducible solution pairs $(A,\psi)$, modulo gauge transformations.
The Seiberg--Witten invariants are then homology classes of $\cal M$,
independent of the metric on $X$.  These invariants prove very useful
in 4-manifold theory.  In particular, Seiberg and Witten give a
`physicist's proof' that the instanton invariants of certain
4-manifolds (namely with $b^+ > 1$, where $b^+$ is the dimension of
the space of self-dual harmonic forms) can be expressed in terms of
the Seiberg--Witten invariants.

From the quantum field theory point of view, the importance of
Seiberg--Witten theory lies in the concept of {\em duality}.  In a
modified version of Yang--Mills theory, called $N=2$ supersymmetric
Yang--Mills theory,  the quantum field theory is described by a scale
parameter $t$ and a complex parameter $u$ (here supersymmetry is
essential).  In the limit $t \to \infty$, the theory is described by
an analytic function $\tau$ of $u$.  If $b^+ (X) > 1$, then $\tau$ is
{\em modular} (in the classical sense) with respect to the action of
$SL(2,\bbz)$.  This means in particular that a theory with parameter
$u$ is related to a theory with parameter $u^{-1}$ in a definite and
known way.  The transformation $u \mapsto u^{-1}$ corresponds to
changing the coupling constant to its inverse.  Hence for the magnetic
monopoles of the theory this represents a {\em duality}
transformation: from {\em electric} with coupling $e$ to {\em
magnetic} with coupling $\tilde{e}$ and vice versa, since Dirac's
quantization condition states that $e \tilde{e} =1$ in suitable
units.  By relating a `strongly coupled' theory to a `weakly coupled'
theory, one can hope to obtain results on the former by performing
perturbative calculations (which are meaningless when coupling is
strong) in the latter.  By inspecting their moduli spaces one is often
able to identify pairs of dually related theories.

\section{String and related theories}
I shall be extremely brief about these theories.  The reason is, apart
from my own obvious ignorance, that they are considerably more
complicated than gauge theories and require much more knowledge not
only of quantum physics but also of algebraic geometry than can
reasonably be dealt with in this workshop.  My aim here is just to
give a taste of some immensely active areas of research in
mathematical physics in recent years where moduli spaces play an
important role.

The gist of string theory is that the fundamental objects under study
are not point-like particles as in gauge field theories but
1-dimensional extended strings.  These strings are really the
microscopic quantum analogues of violin strings: they move in space
and they also vibrate.  The equation  of motion of a free string can be
obtained from an action which is similar to that for a massless free
particle.  In the latter case we have
$$S_0=\int d\tau \eta_{\mu\nu} \frac{dx^\mu}{d \tau} \frac{d x^\nu}{d
\tau} $$
which is just the {\em length} of the `worldline' in spacetime $X$
traced out by the particle as it travels through space.   Here
$\eta_{\mu\nu}$ is the metric on $X$ and $x^\mu$ are the coordinates
of the particle.  For the string the free action is the {\em area} of
the `worldsheet' (with coordinates $\sigma,\tau$) traced out by the
1-dimensional string in spacetime $X$:
$$S_1 = \int d\sigma d\tau \eta^{\alpha\beta} \eta_{\mu\nu}
\partial_\alpha x^\mu \partial_\beta x^\nu, $$
where the indices $\alpha,\beta=0,1$ refer to the worldsheet.  Varying
$S_1$ with respect to $x$ gives simply the 2-dimensional wave equation:
$$ \left( \frac{\partial^2}{\partial \tau^2} -
\frac{\partial^2}{\partial \sigma^2} \right) x^\mu =0.  $$
We see that in this context spacetime coordinates can be regarded as 
fields on the 2-dimensional surface which is the worldsheet.

\begin{figure} [h]
\centerline{\psfig{figure=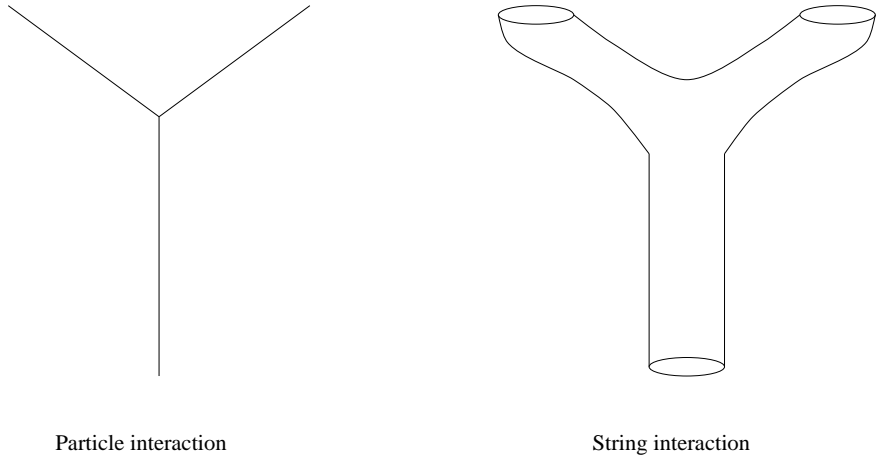,width=0.8\textwidth}}
\caption{Schematic representation of particle and string interactions}
\label{strings}
\end{figure}

\begin{figure} [h]
\centerline{\psfig{figure=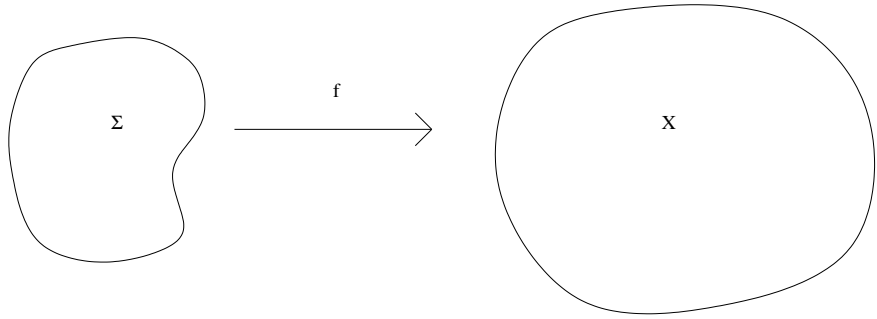,width=0.8\textwidth}}
\caption{Embedding worldsheet into spacetime}
\label{embed}
\end{figure}

Interaction between strings are given by the joining and splitting of
strings so that the resultant worldsheet can be visualized, on
euclideanization, as a Riemann surface $\Sigma$
with a given genus (see Figure
\ref{strings}).  
For example, a hole in $\Sigma$ can be obtained by
one closed string splitting into two and then joining together
again.  In fact, a useful way of looking at string theory is to think
of it as being given by an embedding $f$ of a Riemann surface $\Sigma$
into spacetime $X$ (Figure \ref{embed}).

\subsection{Conformal field theory}
We have written the action $S_1$ for a free string in terms of a
particular parametrization of $\Sigma$, but obviously the physics
ought to be invariant under reparametrization.  The group of
reparametrization on $\Sigma$ is the infinite-dimensional conformal
group, and that is the symmetry group of string theory.

On the other hand, on  a given Riemann surface $\Sigma$ one can
consider certain field theories which have this invariance.  These are
called conformal field theories (CFT) and play important roles in
statistical mechanics and critical phenomena (e.g.\ phase change),
when the theories become independent of the length scale (so that
quantities are defined only up to conformal transformations).

\begin{figure} [h]
\centerline{\psfig{figure=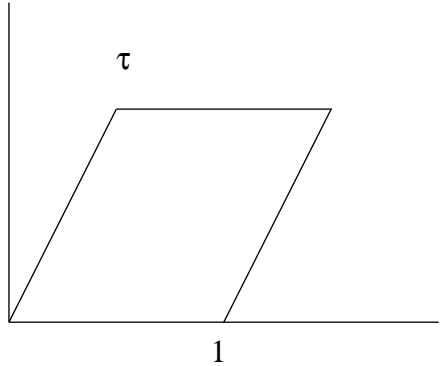,width=0.4\textwidth}}
\caption{A 2-torus represented on the complex plane}
\label{torus}
\end{figure}

The concept of moduli plays an important role in CFT.  In fact, the
original idea of modulus is defined for Riemann surfaces (see talk by
Frances Kirwan).  So a torus $T^2$ has one modulus $\tau$ (see Figure
\ref{torus}).  
The conformal structure of $T^2$ is invariant under the
action of the {\em modular group} $SL(2,\bbz)$ on $\tau$.

CFT are often studied for their own sake, but as far as string
theories are concerned their use lies in the fact that they are the
terms in  a first-quantized, perturbative formulation of string
theory.  Schematically, one can think of string theory as the `sum
over $g$' of CFT on Riemann surfaces of genus $g$.  Unfortunately,
this `summation' has never yet been given a precise meaning.  What
provides some hope that the problem may be tractable is the fact that
the infinite-dimensional integral
$\int e^{-S_1(x)}$ occurring in the path integral formalism can be
reduced to one on the moduli space of the Riemann surface, which is
finite-dimensional. 

\subsection{Various string theories}
Up to now I have been carefully vague about the nature of spacetime
$X$ in string theory.  It turns out that to get a consistent,
first-quantized theory, one needs $X$ to have 26 dimensions!  If we
modify the theory by adding supersymmetry to produce a {\em
superstring} theory, then $\dim X=10$.  However,
this potentially disastrous requirement has been turned to good use to
produce interesting theories in 4 dimensions, as we now briefly sketch.

We shall concentrate on the supersymmetric version as being the more
favoured by string theorists, in that we now assume $\dim X =10$.
Imagine that one can compactify 6 of these 10 dimensions so that
$$ X \cong K \times \bbr^4 $$
with $K$ a compact 6-dimensional space, and moreover that the size of
$K$ is small.  Since the length is an inverse measure of energy, this
means that to observers of low energy (such as us) spacetime will just
look 4-dimensional and the other 6 dimensions are curled up so tight
we cannot see them.  The often-quoted example is that a water pipe
looks like a thin line from a distance.

Not only that, the symmetries of $X$ can be factored into that of
$\bbr^4$ (the usual ones) and that of $K$.  The latter can then be
interpreted as the internal symmetries of Yang--Mills theory.  In
fact, the choice of $K$ is dictated by which gauge symmetry one wants.

There are in all 5 string theories.  A string can be open (homeomorphic
to an interval) or closed (homeomorphic to a circle).  An open string
theory is called Type I.  For closed strings, depending in the
boundary conditions one imposes, one has Type IIA or Type IIB.  If one
combines both the usual and the supersymmetric versions one obtains 
the heterotic
string, with gauge group (after suitable compactification) either $E_8
\times E_8$ or $SO(32)$.  The $E_8 \times E_8$
heterotic string is particularly favoured as being able to include
various Yang--Mills theories which are important in particle physics.

\subsection{$\boldmath M$ Theory}
One can generalize the 1-dimensional strings to higher-dimensional
objects called `membranes'; similarly superstrings to
`supermembranes'.  The study of these last objects have become
particularly fashionable, especially after the introduction of 
something called $M$-theory.

Now supersymmetry can also be made into a local gauge theory which is
then called {\em supergravity}.  It was shown some time ago that in
supergravity, $\dim X \leq 11$, so 11-dimensional supergravity was
studied as being in some sense a unique theory.

$M$-theory is perceived as an 11-dimensional supergravity theory,
where the 11-dimensional manifold $X$ can be variously compactified to
give different superstring theories.  Moreover, solitonic solutions are
found which are supermembranes.  By examining the moduli of these
solutions one can connect pairs of underlying string theories.  For
example, reminiscent of the Seiberg--Witten duality and using the
modular transformations on the modulus $\tau$ of the torus (in one of
the compactifications of $X$), one can connect the two different
versions of the heterotic string.  In fact, by using both
compactification and duality one finds that $M$-theory can give rise
to all the 5 superstring theories mentioned above.  So in some sense,
all the 5 are equivalent and one can imagine that they are just
different perturbative expansions of the same underlying $M$-theory.

Most recently, Maldecena suggested that $M$-theory on compactification
on a particular 5-dimensional manifold (called anti-de Sitter space),
including all its gravitational interactions, may be described by a
(non-gravitational) Yang--Mills theory on the boundary of $X$ which
happens to be 4-dimensional Minkowski space (i.e.\ flat spacetime).
This opens up some new vistas in the field.

Although progress is made in an almost day-to-day basis, we are 
still waiting for a fuller description, perhaps even a
definition, of $M$-theory.  Meanwhile, it has generated a lot of
interest and especially intense study into the various moduli spaces
that occur.

\section{Conclusions}
I have endeavoured to describe a few pysical theories in which moduli
space plays an important role.   However,  I must say that the success
in the reverse direction is more spectacular---using Yang--Mills
moduli spaces (in different specializations) to understand
4-manifolds, following Donaldson, Kronheimer and many others.  At the
beginning I have explained why the success in physics is more
restricted.  Nevertheless, there are many high points:
\begin{enumerate}
\item Self-dual Yang--Mills $\leadsto$ instantons $\leadsto$ vacuum
structure of QCD.
\item Monopole moduli spaces $\leadsto$ identification of pairs of
dual theories in Seiberg--Witten scheme $\leadsto$ hope for
possibility of practical computations in quantum field theory.
\item Classification of conformal field theories $\leadsto$
application of theoretical statistical mechanics.
\item Identifying moduli spaces to connect up the different string
theories $\leadsto$ leading to a unification in 11 dimensions? 
\end{enumerate}

But for lack of time and expertise, I have omitted many other areas of
mathematical physics being actively pursued at present in
which moduli spaces play significant roles.

\section*{References}
The following is only a small selection of articles that I have used
in preparing this talk.  They are in no way even representative.
\begin{enumerate}
\item Self-duality in 4-dimension Riemannian geometry, M.F.\ Atiyah,
N.J.\ Hitchin and I.M.\ Singer, Proc.\ Roy.\ Soc.\ A362 (1978)
425-461.
\item The Seiberg--Witten equations and 4-manifold topology, S.K.\
Donaldson, Bull.\ AMS 33 (1996) 45-70.
\item The first chapter of {\em Superstring theory}, M.B.\ Green,
J.H.\ Schwarz and E.\ Witten, Cambridge University Press, 1987.
\item A layman's guide to $M$-theory, M.J.\ Duff, hep-th/9805177, talk
delivered at the Abdus Salam Memorial Meeting, ICTP, Trieste, November
1997.
\end{enumerate}

\end{document}